\documentstyle[pra,aps]{revtex}

\input epsf

\begin{document}

\draft

\title{Wigner functions, squeezing properties and slow decoherence of 
\\
atomic Schr\"{o}dinger cats}

\author{
{M. G. Benedict$^{1,}$}\cite{MGBemail} 
and 
{A. Czirj\'{a}k$^{1,2,}$}\cite{ACemail}
}

\address{
$^{1}$Department of Theoretical Physics, Attila J\'{o}zsef University, 
\\
H-6720 Szeged, Tisza Lajos krt. 84-86, Hungary
\\
$^{2}$Department of Quantum Physics, University of Ulm, 
D-89069 Ulm, Germany
}

\date{Submitted to Physical Review A: March 26, 1999.}

\maketitle

\begin{abstract}
We consider a class of states in an ensemble of two-level atoms: 
a superposition of two distinct atomic coherent states, which can
be regarded as atomic analogues of the states usually called Schr\"{o}dinger
cat states in quantum optics. According to the relation of the constituents 
we define polar and nonpolar cat states. The properties of these are 
investigated by the aid of the spherical Wigner function. We show that 
nonpolar cat states generally exhibit squeezing, the measure of which 
depends on the separation of the components of the cat, and also on 
the number of the constituent atoms. 
By solving the master equation for the polar cat state
embedded in an external environment, we determine the characteristic times
of decoherence, dissipation and also the characteristic time of a new
parameter, the non-classicality of the state. This latter one is introduced
by the help of the Wigner function, which is used also to visualize the
process. The dependence of the characteristic times on the number of atoms 
of the cat and on the temperature of the environment shows that the 
decoherence of polar cat states is surprisingly slow.
\end{abstract}


%
%
%
%

\section{Introduction}

The question why macroscopic superpositions are not observable in everyday
life has been raised most strikingly by Schr\"{o}dinger in his famous cat
paradox. Recent experiments \cite{MMKW,BrHR}, however, show that at least
mesoscopic superpositions can be observed in quantum-optical systems. In
quantum optics one usually speaks of a Schr\"{o}dinger cat (SC) state if
one has a superposition of two different coherent states of a harmonic
oscillator. In one of the experiments \cite{MMKW} a superposition of two
different coherent states have been created for an ion oscillating in a
harmonic potential. In the other one \cite{BrHR} two coherent states of a
cavity mode were superposed, and also the process of the decoherence between
these states could be followed by monitoring the field with resonant atoms.
The unusual properties of such states have been discussed theoretically
in several publications, see e.g. \cite{YS86,JJ90,SPL91,BMKP92}.

A different type of Schr\"{o}dinger cat like state can be created
in principle in a collection of two-level atoms, as first proposed in 
\cite{CZ94}.
The terminology we may use, is the following: The individual
two-level atoms can be regarded as the ``cells'' of the cat, and the cat is
definitely alive, if all of its cells are alive, i.e. they are in the 
$|+\rangle $ state, and it is definitely dead, if all the cells are in the
ill, $|-\rangle $ state. In the case of $N$ atoms a prototype of a SC like
state is then: 
\begin{equation}
|\Psi _{\text{SC}}\rangle ={\frac{1}{\sqrt{2}}}
(|{+,+,\ldots ,+}\rangle +|{-,-,\ldots ,-}\rangle ),
\label{SC0}
\end{equation}
where each of the terms contain $N$ pluses and $N$ minuses. We shall call
this state the polar cat state, because the two components are in the
farthest possible distance from each other. This state is in the totally
symmetric $N+1$ dimensional subspace of the whole $2^{N}$ dimensional
Hilbert space, and if such states are manipulated by a resonant
electromagnetic field mode with dipole interaction, then the atomic system
will remain in this subspace. This is the arena of the collective
interaction of the atoms and the electromagnetic field, called superradiance
\cite{D54,GH82,BM96}.
In this work we present results concerning the
properties and dynamics of polar cat states (\ref{SC0}), and also of more 
general collective atomic states, the generation of which have also been 
considered recently \cite{AgPS,G98}.

Our approach of discussing the properties of quantum states like 
$|\Psi _{\text{SC}}\rangle $ is based mainly on the method of the Wigner 
function, which is one of the possible quasi-probability distributions. 
It has become a customary tool for investigating quantum states of an
electromagnetic mode oscillator, or an ion oscillating in an appropriate
trapping field \cite{FBLSS97,LMKMIW96}. The method of Wigner function is
much less exploited, however, in the description of atomic states like 
(\ref{SC0}). 
That is why we first summarize the essentials of this method, and
then turn to the determination of the Wigner function for the cat state 
(\ref{SC0}) 
in Section II. Next, in Section III. we consider more general cat
like states, which we call ``nonpolar cats'' and deteremine their squeezing
properties. Finally, in Section IV. we write down and solve the master 
equation for a cat state in an environment with finite temperature. 
We define and determine the dissipation and decoherence times of the system, 
and the characteristic time when the system becomes essentially classical.

%
%
%
%

\section{The Wigner function of the polar cat state}

The $N$-atom dipole interaction with the electromagnetic field is 
equivalent to the dynamics of
a spin of $j=N/2$, and the phase space of the atomic subsystem is the
surface of a sphere of radius $\sqrt{j(j+1)}$, ($\hbar =1$), which is
sometimes called the Bloch sphere. This phase space and quasiprobability
distributions corresponding to various operators 
acting in the $2j+1$ dimensional Hilbert space
have been introduced first by Stratonovich \cite{St}. Similar constructions
have been considered independently by several authors
\cite{ACGT,Ag,G76,VG,SW94}.
We use here the construction and notation introduced
by Agarwal \cite{Ag}. Similarly to the case of oscillator quasidistributions 
\cite{CG,AgW}, 
the quasiprobability functions for angular momentum states
are not unique either. Beyond the natural requirements that the possible
quasiprobability distribution functions have to satisfy, there is a special
property, called the product rule, that distinguishes the most natural
choice among the possible quasiprobability distributions. This rule requires
that the expectation value of a product of two operators could be calculated
by integrating the product of the corresponding quasiprobabilities. This
choice is essentially unique, and in accordance with most authors we call it
the Wigner function for spin $j$. We note that the construction can be
extended to include several values of $j$ 
\cite{FBC98,Wolf}, 
and in the same
spirit Wigner functions can be defined for arbitrary Lie groups \cite{Brif}.
We also note that it is possible to define joint Wigner functions
for atom-field interactions, and then a fully phase space description of
atom-field dynamics can be considered \cite{CB96}. Here we restrict ourselves
to the problem of angular momentum with a fixed value of $j$.

Using the procedure proposed in \cite{Ag} we shortly summarize here the
method of quasiprobability functions in the $2j+1$ dimensional Hilbert
space. One first chooses an operator basis in this space, and the most
straightforward set of operators is the set of the spherical tensor
operators $T_{KQ}$ which transform among others irreducibly under the action
of the rotation operators\cite{BL}. Their explicit expression is:
\begin{eqnarray}
T_{KQ}=\sum_{m=-j}^{j}(-1)^{j-m}(2K+1)^{1/2}\left( 
\begin{array}{ccc}
j & K & j \\ 
-m & Q & m-Q
\end{array}
\right) |j,m\rangle \langle j,m-Q| 
\end{eqnarray}
where 
$\left( 
\begin{array}{ccc}
j & K & j \\ 
-m & Q & m-Q
\end{array}
\right) $ 
is the Wigner $3j$ symbol. 
They form a basis in the sense that any operator 
of the  Hilbert space can be expanded in terms of them and 
they fulfil the Hilbert-Schmidt orthonormality condition 
$\text{Tr} \left( T^{\dagger}_{KQ} T_{K'Q'} \right)=\delta_{KK'}\delta_{QQ'}$.

Introducing the characteristic
matrix of the density operator $\rho $ with respect of this operator basis
as:
\begin{equation}
\varrho_{KQ}=\text{Tr} \left( \rho \,T_{KQ}^{\dagger} \right),  \label{charf}
\end{equation}
the Wigner function of the state $\rho$ is defined as: 
\begin{equation}
W_{\rho }(\theta ,\phi )=\sqrt{\frac{2j+1}{4\pi }}\sum_{K=0}^{2j}
\sum_{Q=-K}^{K}\varrho _{KQ} \, Y_{KQ}(\theta ,\phi ).  \label{WFU}
\end{equation}
The factor in front of the sum ensures normalization.
We note that in a similar way one can associate a Wigner function 
$W_{A}(\theta ,\phi )$ to any operator $A$, by introducing its characteristic
matrix: 
$A_{KQ}=\text{Tr} \left( A\,T_{KQ}^{\dagger} \right)$, 
and then forming the sum as in Eq. (\ref{WFU}). 
It can be easily seen, that this is a very similar procedure according to
which one introduces the quasidistributions of oscillator states and
operators by the help of characteristic functions of the translation operator
basis: $D(\alpha )=\exp (\alpha a^{\dagger }-\alpha ^{*}a)$ \cite{CG,AgW}.
The construction of Eq. (\ref{WFU}) can be shown to satisfy the product rule
mentioned above, giving the following result for the expectation
value of an operator $A$:
\begin{eqnarray}
\text{Tr}
(\rho A)= \sqrt{\frac{4\pi }{2j+1}} \int W_{\rho}(\theta ,\phi )
W_{A}(\theta ,\phi )\sin \theta
\,\text{d}\theta \,\text{d}\phi . 
\end{eqnarray}
For other types of quasidistributions of angular momentum, like the analogs
of the oscillator $P$ and $Q$ functions see \cite{Ag}.

Similarly to the case of the oscillator, the Wigner function allows one to
visualize the properties of the state in question. In the work of Dowling \&
al. \cite{DAS} graphical representations of the Wigner function of the
number, coherent and squeezed atomic states were presented. The Wigner
function of a cat state like (\ref{SC0}) has been considered first in \cite
{BCAS}. 

The characteristic matrix of the state given by (\ref{SC0}) can now be
calculated according to the definition, Eq. (\ref{charf}), taking into
account that the density operator corresponding to
$|\Psi _{\text{SC}}\rangle $ is 
\begin{eqnarray}
\rho _{\text{SC}}=\frac{1}{2}\left( \left| j,j\right\rangle \left\langle
j,j\right| +\left| j,-j\right\rangle \left\langle j,-j\right| +\left|
j,j\right\rangle \left\langle j,-j\right| +\left| j,-j\right\rangle 
\left\langle j,j\right| \right) 
\end{eqnarray}
in the standard basis, with $j=N/2.$ The characteristic matrix has the form: 
\begin{eqnarray}
(\varrho _{\text{SC}})_{K,Q} = \frac{\sqrt{2K+1}}{2}
\left\{  \left( 
\begin{array}{ccc}
j & K & j \\ 
-j & 0 & j
\end{array}
\right) (1+(-1)^{K})\delta _{Q,0} +
\left( 
\begin{array}{ccc}
j & K & j \\ 
-j & 2j & -j
\end{array}
\right) (\delta _{Q,2j}+(-1)^{K}\delta _{Q,-2j})\right\} ,
\end{eqnarray}
and from Eq. (\ref{WFU}) one obtains the following result for the Wigner
function: 
\begin{eqnarray}
W_{\text{SC}}(\theta ,\phi )={\frac{1}{2}}\sqrt{\frac{N+1}{4\pi }} 
\left\{ \sum_{l=0}^{N}{\frac{{\sqrt{2l+1}N!}}{{\sqrt{(N-l)!(N+l+1)!}}}}
[Y_{l0}(\theta )+Y_{l0}(\pi -\theta )] 
+2\sqrt{\frac{{(2N+1)!}}{{4\pi }}}{
\frac{{(\sin \theta )^{N}\cos (N\phi )}}{{2^{N}N!}}}\right\} . \label{WFSC}
\end{eqnarray}
The first term, containing the sums of two spherical harmonics,
corresponds to the individual states $|+,+,\ldots,+\rangle $, and 
$|-,-,\ldots,-\rangle $, while the last term arises from the interference 
term between the ``living'' and ``dead'' parts of (\ref{SC0}) (the last two
terms of the density operator).

Fig. 1 shows the polar diagram of this Wigner function for $N=5$ atoms. The 
two bumps to the ``north'' and ``south'' correspond to the quasiclassical
coherent constituents, while the ripples along the equator -- where the
function takes periodically positive and negative values -- are the result 
of interference between the two kets of Eq. (\ref{SC0}). 
The factor $\cos (N\phi )$ in (\ref{WFSC}) shows that 
the number of negative ``wings''
along the equator is equal to the number of atoms. 

%
\begin{figure}[htbp]
\epsfxsize=3.375in 
\epsfbox{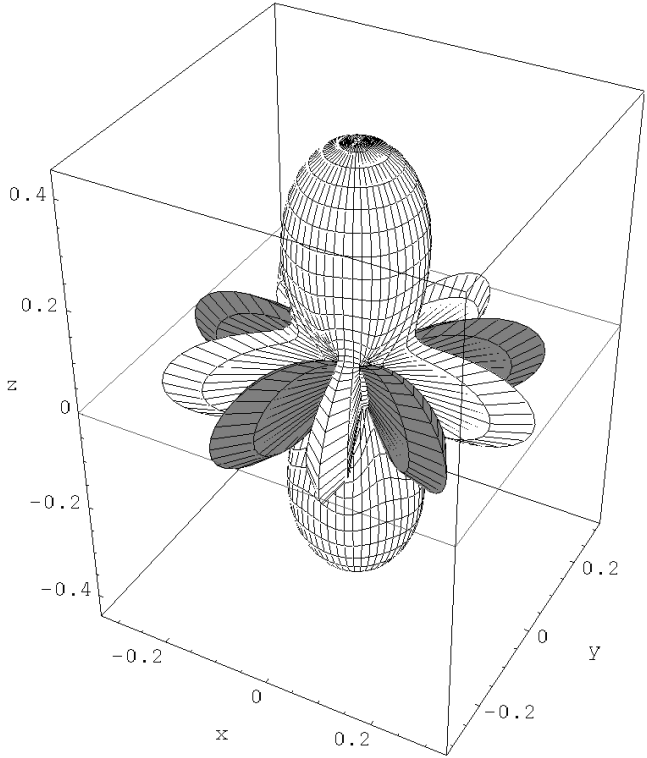} 
\caption{
Wigner function for a polar cat state, Eq. (\ref{WFSC}) for the
case of $N=5$ atoms. The absolute value of the function is measured 
along the
radius in the direction $(\theta ,\phi) $, and the surface is shown in light
where the function is positive and in dark where it takes on negative
values. }
\label{fig:polcat}
\end{figure}
%

%
%
%
%

\section{Nonpolar cat states, and their squeezing properties}

\subsection{Nonpolar cat states}

One can also construct more general SC states by taking the superposition of
any two atomic coherent states. An atomic coherent state (a quasiclassical
state) \cite{ACGT}, $\left| \tau \right\rangle $ is an eigenstate 
with the highest eigenvalue $m=j$
of the component of the angular momentum operator 
`pointing in the direction ${\bf n}$': 
\begin{eqnarray}
\left( {\bf J\cdot n}\right) |\tau \rangle =j|\tau \rangle . 
\end{eqnarray}
The notation $\tau $ refers to a specific parametrization of the unit vector 
${\bf n}$ by its stereographic projection to the complex plane. It is
connected with the polar angle $\beta $ and the azimuth $\alpha $ of the
direction ${\bf n}$ as $\tau =\tan (\beta /2)e^{-i\alpha }$. The
atomic coherent state can be expanded in terms of the eigenstates 
$\left|j,m\right\rangle $ of $J_{z}$ \cite{ACGT}:
\begin{eqnarray}
\nonumber
|\tau \rangle  &=&\left( \frac{1}{1+|\tau |^{2}}\right) ^{j}e^{\tau
J_{+}}\left| j,-j\right\rangle 
= \sum_{m=-j}^{j}
{2j \choose j+m}^{1/2}
\frac{\tau ^{j+m}}{(1+|\tau |^{2})^{j}}\left| j,m\right\rangle  
\nonumber \\
&=&\sum_{m=-j}^{j}
{2j \choose j+m}^{1/2}
\sin ^{j+m}(\beta /2)\cos ^{j-m}(\beta /2)e^{-i(j+m)\alpha }
\left| j,m \right\rangle .   \label{COHS}
\end{eqnarray}
The superposition of two quasiclassical coherent states is given by the ket: 
\begin{equation}
|\Psi _{12}\rangle ={\frac{{|\tau _{1}\rangle +|\tau _{2}\rangle }}
{\sqrt{2(1+\text{Re}\langle \tau _{1}|\tau _{2}\rangle )}}}.  \label{NPSC}
\end{equation}
Recently Agarwal, Puri and Singh \cite{AgPS} and Gerry and Grobe \cite
{G98} have proposed methods to generate such states in a cavity, via a
dispersive interaction with the cavity mode.

%
\begin{figure}[htbp]
\epsfxsize=3.375in 
\epsfbox{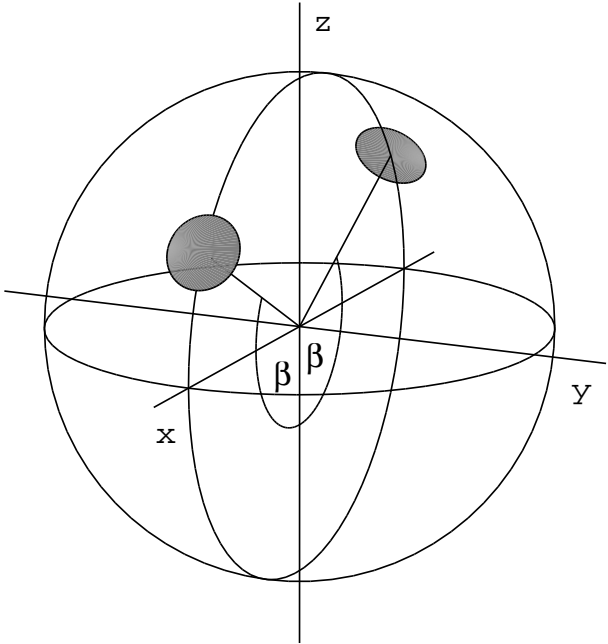} 
\caption{
The phase space scheme of a nonpolar cat state, Eq.(\ref{NPSC}) with
$\tau _{2}=-\tau _{1}$.
}
\label{fig:scheme}
\end{figure}
%

We choose here $\tau _{1}=\tan (\beta /2)$, $\tau _{2}=-\tau _{1}$. Then 
$\beta $ is the polar angle of the classical Bloch vector corresponding to 
the atomic coherent state ${|\tau _{1}\rangle }$ ($\beta $ is measured 
from the south pole), see Fig. 2.  This means that the $x$ component of the
expectation value of the dipole moment in these states is proportional to 
$\pm (N/2)\sin \beta $, respectively, and the $y$ component is zero. Any
other equal weight superposition of two atomic coherent states can be
obtained from this special choice by an appropriate rotation. The polar cat
state of the previous section corresponds to the special case when the two
points are the northern and southern poles of the Bloch sphere. 
If the centres of the two coherent states in question are not
in opposite points of the sphere, then we will call their superposition
as ``nonpolar'' cat states. 

The corresponding
quasiprobability distribution functions still can be explicitly calculated.
For the Wigner function of the cat state $|\Psi _{12}\rangle $ one gets the
following expression: 
\begin{eqnarray} 
W(\theta ,\phi ) =
 \sqrt{\frac{N+1}{4\pi}}
&& \sum_{K=0}^{2j}\sum_{Q=-K}^{K}{\frac{\sqrt{2K+1}{(2j)!}}
{{2(1+(\cos \beta )^{2j})}}}{\ }
 \sum_{m=-j}^{j}{\frac{
(-1)^{j-Q-m}+(-1)^{3j+m}+(-1)^{2j}+(-1)^{2j-Q}}{\sqrt{
(j+m)!(j-m)!(j+Q+m)!(j-Q-m)!}}} \hfill  
 \nonumber \\
&& \times \left( 
\begin{array}{ccc}
j & \ K & \ j \\ 
-m-Q & Q & m
\end{array}
\right) (\sin {\beta /2})^{2(j+m)+Q}(\cos {\beta /2})^{2(j-m)-Q}\
Y_{KQ}(\theta ,\phi ).  \label{WFNP}
\end{eqnarray}
We present polar plots of this Wigner function in Fig. 3. for $N=5$ atoms 
and for several values of $\beta$.

%
\begin{figure}[htbp]
\epsfxsize=7.in
\epsfbox{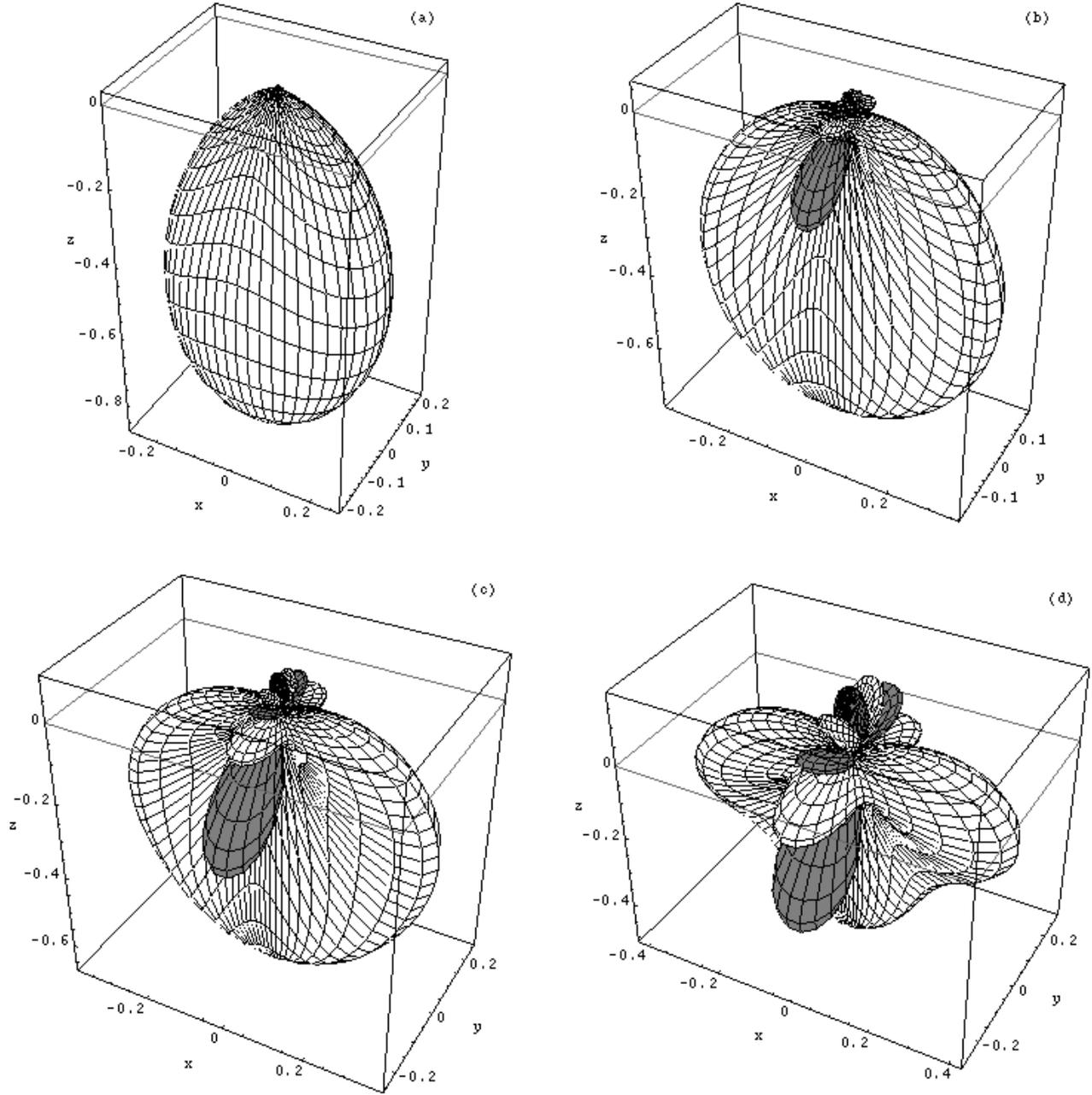}
\caption{
Wigner functions for the state $|\Psi_{12}\rangle$ for $N=5$ atoms,
and for several different values of $\beta$: (a) $\beta=20^{\circ}$, (b) 
$\beta=45^{\circ}$, (c) $\beta=55^{\circ}$, (d) $\beta=70^{\circ}$. For
smaller values of $\beta$ the state goes over into a single coherent state,
and then it has essentially only one positive lobe. This graphical
presentation shows qualitatively that the $y$ component of the dipole moment
is squeezed, the maximal value of the squeezing in the present case ($N=5$)
comes about $\beta=43^{\circ}$. 
}
\label{fig:genpolcat}
\end{figure}
%

For small $\beta $ values, the interference is weak and the maximum of the
Wigner function is around $\theta =0$. For larger $\beta $-s the function
has two maxima around $\theta =\pm \beta $, and the interference gets more
pronounced. When $\beta =\pi /2$, the two maxima corresponding to the
individual coherent states point in the $x$ and $-x$ directions,
respectively. In this case we get back the Wigner function of the SC state
of Eq. (\ref{SC0}), rotated around the $y$ axis by $\pi /2$.

\subsection{Squeezing properties}

The expectation values of the dipole operators $J_{x}$ and $J_{y}$ are zero
in the state (\ref{NPSC}) with $\tau _{1}=\tan (\beta /2)$, $\tau _{2}=-\tau
_{1}$, which is a consequence of the mirror symmetry of this state with
respect of both the $x-z$ and the $y-z$ planes. As it is known, the
variances of the dipole operators, $J_{x}$ and $J_{y}$ are equal to each
other in an atomic coherent state: 
\begin{equation}
(\Delta J_{x})^{2}|_{\left| \tau \right\rangle }=(\Delta J_{y})^{2}|_{\left|
\tau \right\rangle }=j/2.  \label{VCOH}
\end{equation}
In order to calculate the variances in the state (\ref{NPSC}), one can use 
directly expansion (\ref{COHS}) and the known matrix elements of 
$J_{x}$ and $J_{y}$, but the summations that occur
are rather cumbersome to evaluate.  A more effective procedure is to apply
the method of generating functions \cite{ACGT}.  All the necessary
expectation values in a cat state can be calculated by the formula : 
\begin{eqnarray}
\left[ \left( \frac{\partial }{\partial \xi }\right) ^{a}\left( \frac{
\partial }{\partial \eta }\right) ^{b}\left( \frac{\partial }{\partial \zeta 
}\right) ^{c}X_{A}\right] _{\xi =\eta =\zeta =0}=\left\langle \tau
_{1}\right| J_{-}^{a}J_{z}^{b}J_{+}^{c}\left| \tau _{2}\right\rangle ,
\end{eqnarray}
where 
\begin{eqnarray}
\nonumber
X_{A}(\xi ,\eta ,\zeta ) & \equiv & \left\langle \tau _{1}\right| e^{\xi
J_{-}}e^{\eta J_{z}}e^{\zeta J_{+}}\left| \tau _{2}\right\rangle 
=\frac{
\left\langle -j\right| e^{(\tau _{1}^{*}+\xi )J_{-}}e^{\eta J_{z}}e^{(\tau
_{2}+\zeta )J_{+}}\left| -j\right\rangle }{\{(1+|\tau _{1}|^{2})(1+|\tau
_{2}|^{2})\}^{j}} \nonumber  \\ 
&=&\frac{\{e^{-\eta /2}+e^{\eta /2}(\tau _{1}^{*}+\xi )(\tau _{2}+\zeta )\}}
{\{(1+|\tau _{1}|^{2})(1+|\tau _{2}|^{2})\}^{j}}^{2j}
\end{eqnarray}
is the (antinormally ordered) generating function. 

Inserting the necessary operators, we obtain the following expressions 
for the variances in the state given by (\ref{NPSC}): 
\begin{eqnarray}
(\Delta J_{x})^{2}={\frac{j}{2}}\left( 1+{\frac{(2j-1)\sin ^{2}\beta }{{
1+(\cos \beta )^{2j}}}}\right) ,
\end{eqnarray}
\begin{eqnarray}
(\Delta J_{y})^{2}={\frac{j}{2}}\left( 1-{\frac{(2j-1)(\cos \beta
)^{2j-2}\sin ^{2}{\beta }}{{1+(\cos \beta )^{2j}}}}\right) .
\label{YC}
\end{eqnarray}
Comparing these results with Eq. (\ref{VCOH}), we see, that except for some
special cases the $J_{y}$ quadrature is squeezed while the $J_{x}$
quadrature is stretched in this state. The reason of this asymmetry lies in
the fact, of course, that in the superposition (\ref{NPSC}) we have chosen
states that are both centered in points lying in the $x-z$ plane.

One of the exceptional cases that is not squeezed is, if there is only one
atom: $j=1/2.$ As it is easily seen, for $j=1/2$ any state in the
two-dimensional Hilbert space is a coherent state, and therefore it does not
show squeezing. The two other exceptions are $\beta =0$ for any $j$, because
then the two coherent states coincide, and $\beta =\pi /2$, which is the
rotated version of the polar cat state.

Writing Eq.(\ref{YC}) in the form $(\Delta J_{y})^{2}={j}(1-{\cal{S}})/2$, 
we can define the quantity ${\cal{S}}$ as the measure of squeezing. 
Analysis shows that if $N$ is large enough, then 
the maximum value of ${\cal{S}}$ is $0.56$ and it is achieved around 
$\beta _{m}=1.6/\sqrt{N}$. Figure 4 shows the dependence of ${\cal{S}}$ 
on $\beta $ for several values of $N=2j$. 

%
\begin{figure}[htbp]
\epsfxsize=3.375in
\epsfbox{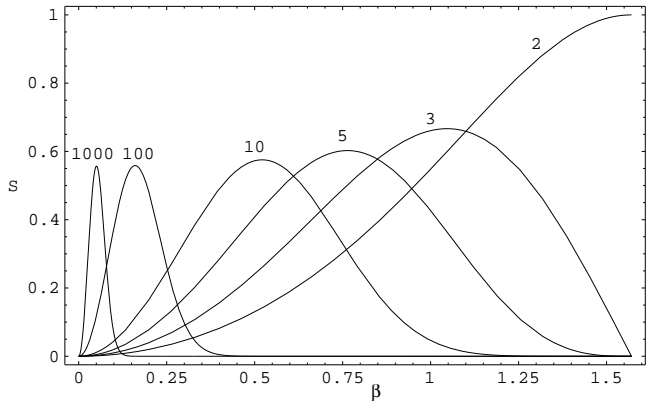}
\caption{
The $\beta$ dependence of the quantity ${\cal{S}}$ in $(\Delta J_y)^2 =
j(1-{\cal{S}}(\beta,j))/2$, for several values of $N=2j$. ${\cal{S}}$ can be
considered as the measure of squeezing for a cat state consisting of two
atomic coherent states separated by the central angle $2\beta$ on the Bloch
sphere.}
\label{fig:sq}
\end{figure}
%

%
%
%
%

\section{Decoherence and dissipation}

As we mentioned in the Introduction, there have already been 
realizable methods proposed for the experimental generation of atomic 
SC states in a collection of two-level atoms \cite{AgPS,G98}. However, 
such an atomic ensemble can never be perfectly isolated from the surrounding 
environment. Further, any observation of these states necessarily leads 
to the interaction of the atomic system with a measuring apparatus.
In both of these cases the atomic system interacts with a system containing
a large number of degrees of freedom. A possible and successful approach 
to this problem \cite{Zu} considers that the static environment continously 
influences the dynamics of the atomic subsystem, which besides exchanging 
energy with the environment loses the coherence of its quantum 
superpositions and evolves into a classical statistical mixture.

In this section we investigate the decoherence and dissipation of the atomic
Schr\"{o}dinger cat states embedded in an environment with many degrees of 
freedom, by writing down the master equation for the reduced 
density operator of the atomic subsystem. 
We provide the solution for the polar cat states (\ref{SC0}).

\subsection{Model and solution}

We couple our ensemble of two-level atoms to the environment  
which is supposed to be a multimode electromagnetic radiation
with photon annihilation and creation operators $a_k$ and $a_k^\dagger$.
Then the interacting system can be described by the following well known 
model Hamiltonian which considers dipole interaction and uses the rotating 
wave approximation: 
\begin{equation}
H=\omega _{\text{a}}J_{z}+ \hbar \sum_{k}\omega _{k}a_{k}^{\dagger
}a_{k}+ \sum_{k}g_{k}\left( a_{k}^{\dagger }J_{-}+a_{k}J_{+}\right) ,
\label{model_ham}
\end{equation}
where $\omega_{\text{a}}$ is the transition frequency between the two atomic 
energy levels, the $\omega_k$ denote the frequencies of the modes of the 
environment and $g_k$ are the coupling constants.
If we suppose the environment to be in thermal
equilibrium at temperature $T$,
then the time evolution of the atomic subsystem is determined by a master 
equation for its reduced density operator $\rho(t)$ 
\cite{ASp,WM}:
\begin{eqnarray}
\hbar^2
\frac{\text{d}\rho (t) }{\text{d}t} = -\frac{\gamma }{2}
\ (\langle n \rangle+1)\ (J_{+}J_{-}\rho (t)
+\rho (t) J_{+}J_{-}-2J_{-}\rho (t) J_{+})
 -\frac{\gamma }{2}
\ \langle n \rangle\ (J_{-}J_{+}\rho (t)
+\rho (t) J_{-}J_{+}-2J_{+}\rho (t) J_{-})  \label{mastereq}
\end{eqnarray}
which involves the usual Born-Markov approximation and is written 
in the interaction picture. Here  
$\langle n \rangle=\left( \exp \left( \hbar \omega _{\text{a}}/
(k_{\text{B}} T)\right) -1\right) ^{-1}$ 
is the mean number of photons in the environment and
$\gamma = (g(\omega_{\text{a}})\sigma(\omega_{\text{a}}))^2$ denotes the 
damping rate, where $\sigma$ is the mode density of the environment.

Eq. (\ref{mastereq}) can be obtained also in a somewhat different 
context, as described in \cite{BSH}. Then one assumes the atomic subsystem
to be placed in a resonant cavity with low quality mirrors causing
the damping of the cavity mode at a rate $\kappa$. Under certain reasonable
assumptions one can get Eq. (\ref{mastereq}) with 
$\gamma = 2 \, g(\omega_{\text{a}})^2/\kappa$.

From Eq. (\ref{mastereq}) 
one can easily deduce the following equations for the matrix elements of the
density operator $\rho _{m,l}(t)\equiv \left\langle j,m\left| \rho
(t)\right| j,l \right\rangle $: 
\begin{eqnarray}
\frac{\text{d}\rho _{m,l}(t)}{\text{d}t}& = &
-\frac{\gamma}{2} \, [ \, \langle n \rangle 
\left( 2j(j+1)-m(m+1)-l(l+1)\right)  
 +(\langle n \rangle+1)\left( 2j(j+1)-m(m-1)-l(l-1) 
\right) \, ] \, \rho _{m,l} (t) 
\nonumber \\  
& & + \,\gamma \, \langle n \rangle\, \sqrt{\left(
j(j+1)-m(m-1)\right) \left( j(j+1)-l(l-1)\right) }\ \rho
_{m-1,l-1} (t)  
 \nonumber \\
& & + \,\gamma \, (\langle n \rangle+1)\, \sqrt{\left( j(j+1)-m(m+1)\right) 
\left( j(j+1)-l(l+1)\right) }\ \rho _{m+1,l+1} (t) . \label{dens_mtrx_dyn}
\end{eqnarray}
Thus the time evolution of a particular density matrix element is coupled 
only to the two neighbouring elements in the corresponding diagonal for 
$\langle n \rangle > 0$, 
and only to the neighbour with larger index at zero temperature.  

In the case of a polar
cat state (consisting of $N=2j$ atoms), the elements of the density
matrix have zero initial values except for $\rho _{-j,-j},\
\rho _{j,j},\ \rho _{-j,j}$ and $\rho _{j,-j}\ (=\rho _{-j,j}^{*}).$ This
implies that the density matrix elements, except for those in the main
diagonal and for $\rho _{-j,j}$ and $\rho _{j,-j}$, remain identically zero
for any time. Setting $\gamma =1$ (i.e. the time unit is $1/\gamma $) the
equations for the elements in the main diagonal of the density matrix are
the following: 
\begin{eqnarray}
\frac{\text{d}\rho _{m,m} (t)}{\text{d}t} &=& 
-\left[ \langle n \rangle\, \left( j(j+1)-m(m+1)\right) 
+(\langle n \rangle+1)\left( j(j+1)-m(m-1)\right)
\right] \ \rho _{m,m} (t) \nonumber \\
& & + \  \langle n \rangle \, \left( j(j+1)-m(m-1)\right) \
\rho _{m-1,m-1} (t)  
 \nonumber \\
& & + \  (\langle n \rangle+1)\, \left( j(j+1)-m(m+1)\right) 
\ \rho _{m+1,m+1} (t) \label{main_diag_dyn}
\end{eqnarray}
with the initial values $\rho _{m,m}(t=0)=\frac{1}{2}(\delta _{m,j}+\delta
_{m,-j})$ (cf. equation (\ref{SC0})). The dynamics of $\rho _{-j,j}$
is governed by the particularly simple equation: 
\begin{equation}
\frac{\text{d}\rho _{-j,j} (t)}{\text{d}t}=
-j(2 \langle n \rangle + 1)\rho _{-j,j} (t) ,
\label{decoh_dyn}
\end{equation}
yielding immediately the following solution with the initial value
$\rho_{-j,j}(0)=1/2$ corresponding to the polar cat state:
\begin{equation}
\rho _{-j,j}(t)=\frac{1}{2}\exp \left( -j(2 \langle n \rangle + 1)t\right).  
\label{decoh}
\end{equation}

As expected, the stationary solution of Eq.~(\ref{main_diag_dyn}) is the 
Boltzmann distribution of the stationary values $\bar{\rho}_{m,m}$:
\begin{eqnarray}
\bar{\rho}_{m,m} & = &
\exp \left( -(m+j) \frac{\hbar \omega_{\text{a}}}{k_{\text{B}} T}  \right) 
\frac{ 1-\exp \left(-\hbar \omega_{\text{a}}/(k_{\text{B}} T) \right) }{
1-\exp \left(- (2j+1) \hbar \omega_{\text{a}}/(k_{\text{B}} T) \right) }
= \frac{
\left( \langle n \rangle / (\langle n \rangle + 1 ) \right)^{m+j}
}{ (\langle n \rangle+1)\ \left(1-\left(\langle n \rangle / 
(\langle n \rangle + 1)\right)^{2 j+1} \right)}.
\label{stac}
\end{eqnarray}
Approximate analytical time dependent solutions of Eq. (\ref{main_diag_dyn}) 
can be found especially for the case of superradiance, when 
$\rho_{j,j}(0)=1$ in \cite{DG}, see also \cite{BM96} and references therein. 
For the initial conditions corresponding to the polar cat state, the time 
dependent solution of equations (\ref{main_diag_dyn}) at zero temperature 
($\langle n \rangle=0$) can be obtained by the following recursive 
integration: 
\begin{eqnarray}
\rho _{j,j}(t) &=&\frac{1}{2}\exp \left( -2jt\right) , \nonumber
\\ \label{recursion}
\\
\rho _{m,m}(t) &=&\exp \left( -b_{m}t\right) \left( \frac{1}{2}\ \delta
_{m,-j}+b_{m+1}\int\limits_{0}^{t}\exp \left( b_{m}t^{\prime }\right) \ \rho
_{m+1,m+1}(t^{\prime })\ \text{d}t^{\prime }\right) ,\hspace{1cm}-j\leq m<j
\nonumber
\end{eqnarray}
where $b_{m}=j(j+1)-m(m-1)$. These equations show rather explicitly, how does
the initial excitation cascade down to the zero temperature stationary state.

For non-zero temperatures ($\langle n \rangle>0$) we have solved equations 
(\ref{main_diag_dyn}) numerically. We are going to analyze the solutions 
in the next subsection.

\subsection{Characteristic times}

\begin{figure}[htbp]
\epsfxsize=7.in
\epsfbox{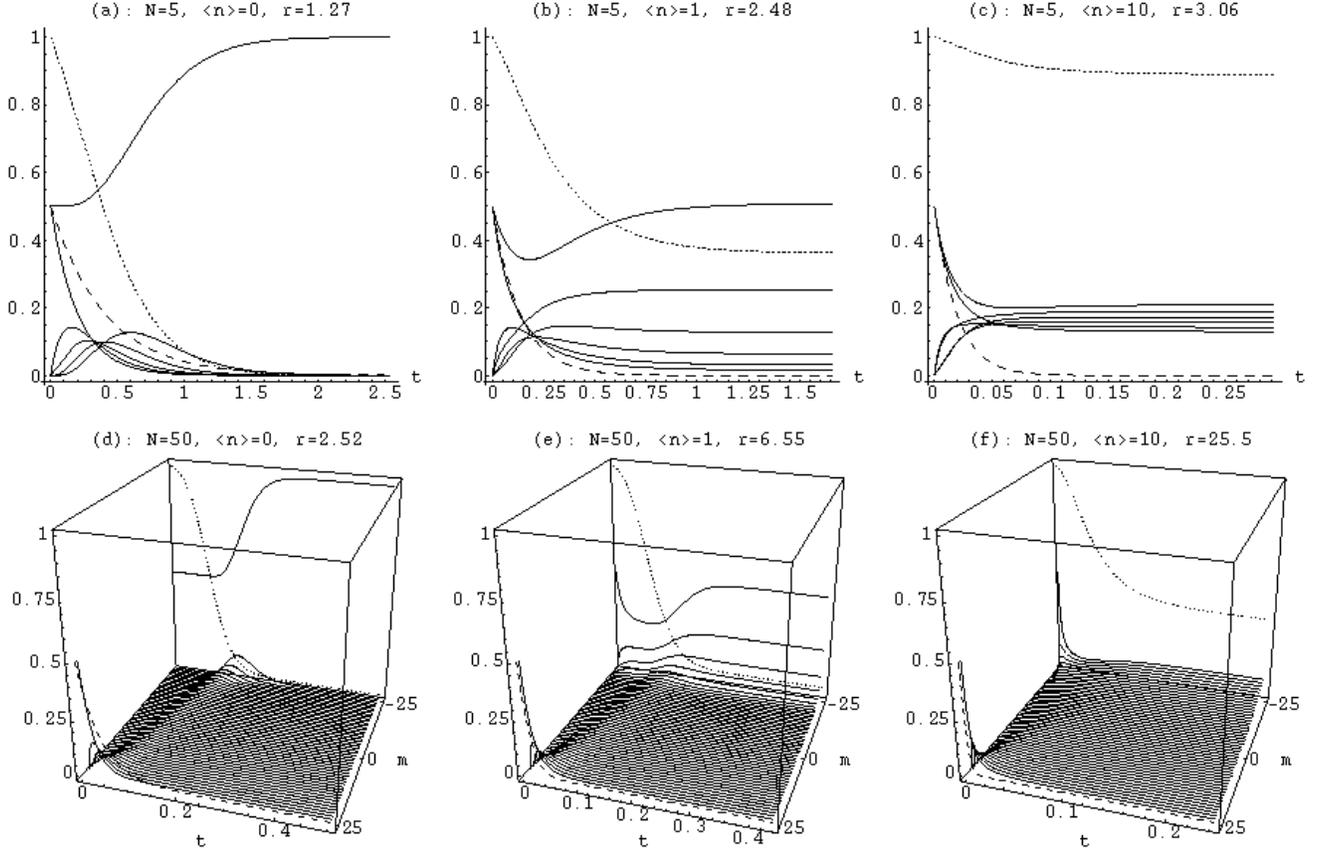}
\caption{
Plots of the density matrix elements $\rho_{m,m} (t)$, 
$m=-j,-j+1,\ldots,j$ (solid lines) and $\rho_{-j,j} (t)$ (dashed line) 
versus time (the time unit is $1/\gamma$). 
Plots are given for $N=5$ atoms: (a, b, c), and $N=50$ atoms: (d, e, f), 
for $\langle n \rangle=0$ (zero temperature): (a, d), 
for $\langle n \rangle=1$: (b, e),
and for $\langle n \rangle=10$: (c, f). 
The solid lines may be identified as follows: 
in (a) the smaller $m$ is the later the corresponding 
$\rho_{m,m}$ has its maximal value; in (b) and (c) 
the stationary values of the $\rho_{m,m}$-s 
follow the Boltzmann distribution, see Eq.~(\ref{stac}).
In (d, e, f) the $\rho_{m,m}$-s follow each other along the axis 
labeled by the $m$.
The dotted lines (starting from 1 at $t=0$) show the normalized energy 
of the atomic subsystem $1+E(t)/(-j \hbar \omega_{\text{a}})$, 
see Eq.~(\ref{energy}).  
The number $r=t_{\text{diss}}/t_{\text{dec}}$ 
is the ratio of the characteristic time of dissipation
to the characteristic time of decoherence.
}
\label{fig:dmplot}
\end{figure}

Figure \ref{fig:dmplot} shows the time evolution of the relevant density 
matrix elements $\rho_{m,m} (t)$, $m=-j,-j+1,\ldots,j$ 
(solid lines) and $\rho_{-j,j} (t)$ (dashed line),
in the case of initial polar cat states consisting of 5 and 50 atoms, 
for $\langle n \rangle=0,\ 1,\ 10$. 

The actual value of  
$\rho_{-j,j}(t)$ characterizes the coherence of the corresponding state, 
since $\rho_{-j,j}$ and $\rho_{j,-j}$ ($=\rho_{-j,j}^*$) are the only
nonzero matrix elements outside the main diagonal. Their
exponential decay (cf. Eq.~(\ref{decoh})) is the decoherence, shown by the 
dashed lines in the plots of Fig. \ref{fig:dmplot}.
Thus it is reasonable to define the characteristic time of the decoherence 
by 
\begin{eqnarray}
t_{\text{dec}} = \frac{2}{N(2 \langle n \rangle + 1)}, 
\label{tdec}
\end{eqnarray}
implying 
$\rho_{-j,j} (t_{\text{dec}})=\rho_{-j,j}(0)/e$. 

In contrast to the simple time dependence of $\rho_{-j,j}(t)$, 
the dynamics of the main diagonal elements $\rho_{m,m} (t)$ depend 
on the actual value of $\langle n \rangle$ and $N$ rather sensitively.
The zero temperature cases, Fig. \ref{fig:dmplot} (a) and (d), clearly show 
the initial excitation, contained in $\rho_{j,j} (0)= 1/2$, cascading 
down to $\rho_{-j,-j} (\infty)=1$ as given by Eqs. (\ref{recursion}). 
At nonzero temperatures ($\langle n \rangle > 0$) 
the time evolution of the $\rho_{m,m} (t)$-s is more complicated because of
the coupling to both neighbours, cf. Eq. (\ref{dens_mtrx_dyn}).

More information can be extracted from the time evolution of the 
$\rho_{m,m} (t)$-s by calculating the energy of the atomic subsystem 
as the function of time:
\begin{eqnarray}
E(t)  \equiv  \langle \omega_{\text{a}} \, J_z \rangle (t) 
 =  \omega_{\text{a}} \text{Tr} \left( \rho (t) \, J_z \right)
 =  \hbar \omega_{\text{a}} \sum_{m=-j}^{j} \, m \, \rho_{m,m} (t).
\label{energy}
\end{eqnarray}

%
%

The process of dissipation (i.e. the change of the energy of the 
atomic subsystem in time) can be very easily followed by studying $E(t)$. 
This function, normalized to the zero temperature stationary energy and 
shifted to vary from 1 to its stationary value: 
$1+E(t)/(-j \hbar \omega_{\text{a}})$,
is shown in the plots of Fig.\ \ref{fig:dmplot} by the dotted lines.
Since its asymptotic behavior is exponential-like, it is reasonable to 
define the characteristic time of dissipation $t_{\text{diss}}$ by 
requiring
\begin{equation}
|E(t_{\text{diss}})-E(\infty)|= |E(0)-E(\infty)|/e.
\end{equation}
In order to ensure that $E(t)$ achieves its stationary
value with a good accuracy in the plots of Fig. \ref{fig:dmplot}   
we have set the time range to $5 \, t_{\text{diss}}$. 
It is seen that the value of $r=t_{\text{diss}}/t_{\text{dec}}$ grows 
with both the temperature and the number of atoms.
A more detailed analysis of this question follows later in this section.

The initial state of the process, the polar cat state, 
has sharply non-classical features. On the other hand, 
at non-zero temperature the final stationary 
state of the present model is a thermal state, 
which is classical in its nature. (At zero temperature the stationary 
state is also non-classical, since it is the state $|j,-j\rangle$.)
It is natural to ask, when does the transition from the non-classical to
the classical stage occur? What is a good measure of non-classicality
reflecting the change of non-classical nature of the corresponding
state? 

The spherical Wigner function (\ref{WFU}) provides a good answer to
both of these questions. 
Quantum states are generally considered essentially
non-classical if the corresponding Wigner function takes on also negative 
values. 
Therefore to answer the second question, 
for the measure of the degree of non-classicality we propose to use the 
quantity $\nu = 1 - (I_+ - I_-)/(I_+ + I_-)$, where $I_+$ is the integral 
of the Wigner function over those domains where it is positive while $I_-$
is the absolute value of the integral of the Wigner function over
the domains where it is negative. Since the integral of the 
Wigner function over the sphere is 1, $I_+ - I_- = 1$, thus 
$\nu = 2 I_- /(2 I_- + 1)$, and it is easy to see that $0 \leq \nu < 1$.
According to this definition, the bigger is the value of $\nu$, 
the more non-classical is the state, and for all classical states 
one has $\nu=0$. 

Regarding now the first question, namely for how long is the state of the 
atomic system non-classical, 
we introduce a third kind of characteristic time $t_{\text{ncl}}$.
We define $t_{\text{ncl}}$ to be the time instant when the corresponding 
spherical Wigner function becomes non-negative everywhere on the sphere,
i.e. $\nu$ becomes 0. 
We will return to this question in connection with the 
time evolution of the Wigner function, which we will present in 
the next subsection in more detail.

\begin{figure}[htbp]
\epsfxsize=3.375in
\epsfbox{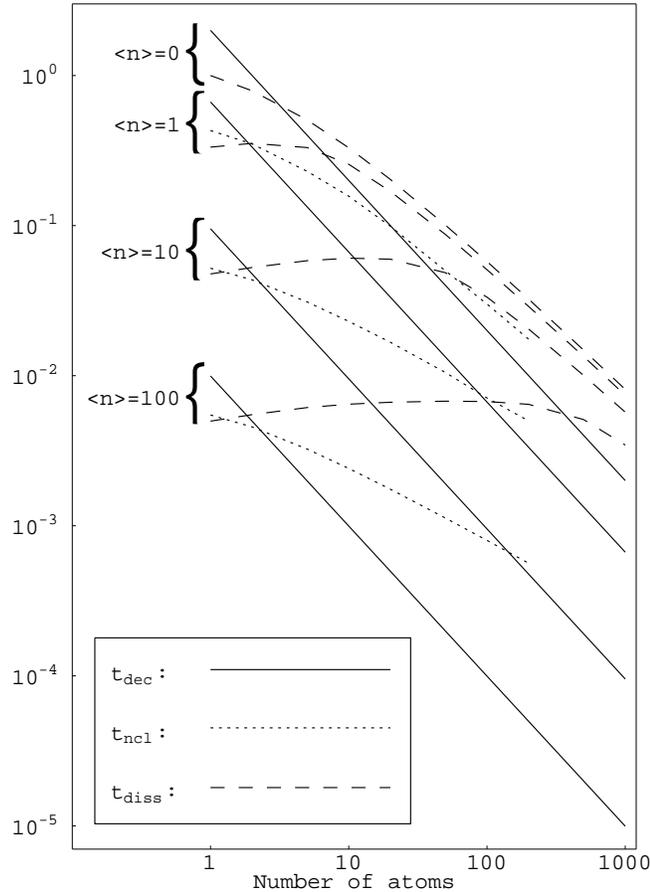}
\caption{
Plots of the characteristic times of 
decoherence, $t_{\text{dec}}$ (solid line), 
dissipation, $t_{\text{diss}}$ (dashed line) and 
non-classicality, $t_{\text{ncl}}$ (dotted line)
versus the number of atoms, on a log-log scale.
The uppermost solid and dashed line are for $\langle n \rangle=0$ 
(there is no plot for $t_{\text{ncl}}$ at zero temperature, 
since the state stays non-classical),
while the subsequent groups of the three kinds of lines, one under the
other, are for $\langle n \rangle=1,\ 10,\ 100$, respectively.}
\label{fig:cht}
\end{figure}

Based on the information provided by the three kinds of 
characteristic times, we consider here the dependence of the 
process on the number of atoms and on the temperature. 
In Figure \ref{fig:cht} we plot $t_{\text{diss}}$ (dashed line), 
$t_{\text{dec}}$ (solid line) and $t_{\text{ncl}}$ (dotted line) 
as the function of the number of constituent atoms of the polar cat $N$, 
for several temperatures, on a log-log scale.

%
%

It is seen that the characteristic time of decoherence $t_{\text{dec}}$
is inversely proportional to the number of
atoms (the straight solid lines in Fig. \ref{fig:cht}), according to 
the definition (\ref{tdec}).
Compared to this, the characteristic time of non-classicality  
$t_{\text{ncl}}$ decreases less rapidly with increasing number of atoms. 
The characteristic time of dissipation $t_{\text{diss}}$ first slightly 
increases at non-zero temperature then it achieves a maximum which depends 
on $\langle n \rangle$ and finally it decreases nearly inversely 
proportional to the number of atoms.
The values of $t_{\text{diss}}$ at different temperatures seem to converge 
slowly beyond a certain number of atoms. 

It seems however rather surprising that the ratio
$t_{\text{diss}}/t_{\text{dec}}$ is not as large as such a quantity is 
usually expected to be \cite{Zu,large}:
in the case of $N=1000$ it is 4.04 for $\langle n \rangle=0$, and
it is still just around 350 for $\langle n \rangle=100$. 
(Note that $\langle n \rangle=100$
corresponds to a temperature of 250~K in the case of typical 
experiments \cite{BrHR}.)
The ratio $t_{\text{diss}}/t_{\text{dec}}$ seems not even to vary 
considerably with increasing $N$ beyond the maximum of $t_{\text{diss}}$ 
mentioned above.
Thus the process of decoherence is extremely slow in the case of a polar 
cat state which is coupled to the environment by an interaction leading to
the master equation (\ref{mastereq}). 

Similar effects have already been reported  for other physical systems
earlier \cite{similar}.
In a recent work Braun, Braun and Haake \cite{BBH} 
investigated the decoherence of an atomic SC state 
$| \tau_1 \rangle + | \tau_2 \rangle$ based on Eq. (\ref{mastereq}) 
for zero temperature.
By evaluating a certain quantity characterising the decoherence rate
at the {\it initial} time, and applying a semiclassical procedure for
finite times
they concluded that for atomic SC states with $\tau_1 \tau_2^* = 1$ 
the decoherence slows down.

Our initial state, the polar cat state fulfils the former condition. 
The results presented in Fig. \ref{fig:cht} derive from the 
solution of the master equation for the whole process.
They are in agreement with the statements of Ref.\cite{BBH}, where the 
initial stage of the decoherence is analyzed for the case of zero 
temperture.

\subsection{Wigner functions}

We illustrate now the process of decoherence and dissipation 
using the spherical Wigner function (\ref{WFU}).

%
\begin{figure}[htbp]
\epsfxsize=7.in
\epsfbox{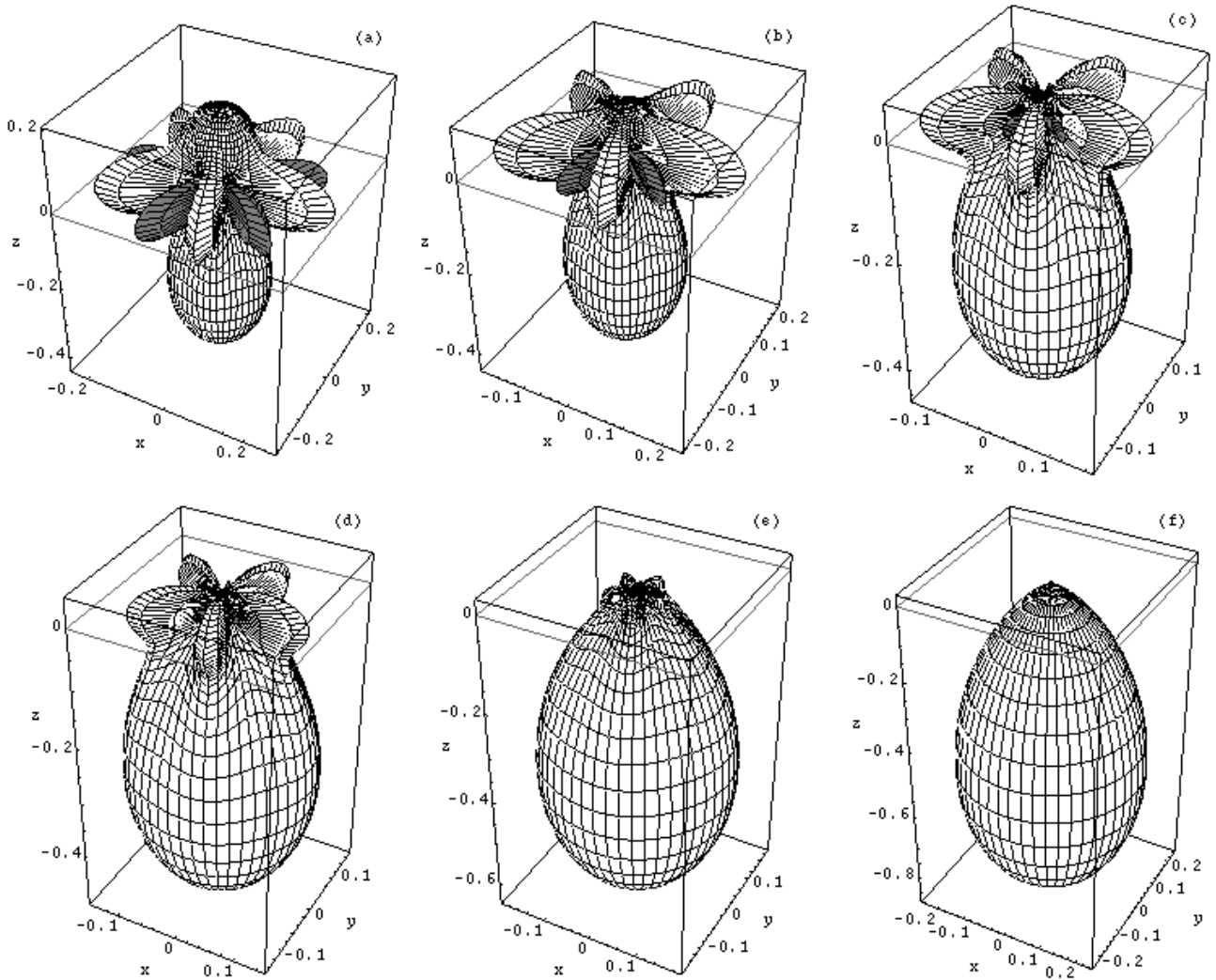}
\caption{
Polar plots of the temporal change of the Wigner function representing the 
decoherence and
dissipation of the initial polar cat state (\ref{SC0}), made of 5 atoms and 
shown in Fig.~\ref{fig:polcat}, at zero temperature.
The dynamics of the corresponding density matrix elements is shown in 
Fig.~\ref{fig:dmplot}(a). The time instants are the following (in units of 
$1/\gamma$): (a) 0.1, (b) 0.2, (c) 0.4 ($=t_{\text{dec}}$), 
(d) 0.506 ($=t_{\text{diss}}$), (e) 0.8, (f) 2.5. 
}
\label{fig:wfN5n0}
\end{figure}
%

In order to obtain its time dependence 
we have to calculate first the characteristic matrix
$\varrho _{K,Q}(t)=\text{Tr} \left( \rho (t)T_{K,Q}^\dagger \right) $ 
from the matrix elements $\rho_{m,l}(t)$ according to 
\begin{equation} 
\varrho _{K,Q}(t)=\sqrt{2K+1}\sum_{m=-j}^{j}(-1)^{j-m}\left( 
\begin{array}{ccc}
j & K & j \\ 
-m & Q & m-Q
\end{array}
\right) \rho _{m,m-Q}(t).
\label{rhotrf}
\end{equation}
From Eq. (\ref{rhotrf}) it can be seen that only 
$\varrho _{K,0}$ ($K=0,1,\ldots,N$) and 
$\varrho _{N,N}=(-1)^{N} (\varrho _{N,-N})^{*}$ are nonzero.
This fact (which is due to the initial conditions specified by the 
polar cat state)
ensures that the azimuthal dependence of the spherical Wigner function 
is determined only by the real part of the spherical harmonic $Y_{N,N} 
(\theta,\phi)$ which is proportional to $\cos (N \phi)$. 
Therefore the Wigner function keeps its initial azimuthal symmetry 
during the whole process. 
Further, since $\varrho _{N,N} (t) =(-1)^{N} \rho _{j,-j} (t)$,
the azimuthal modulation of the spherical Wigner function explicitly
shows the degree of the coherence of the actual state.

\begin{figure}[htbp]
\epsfxsize=7.in
\epsfbox{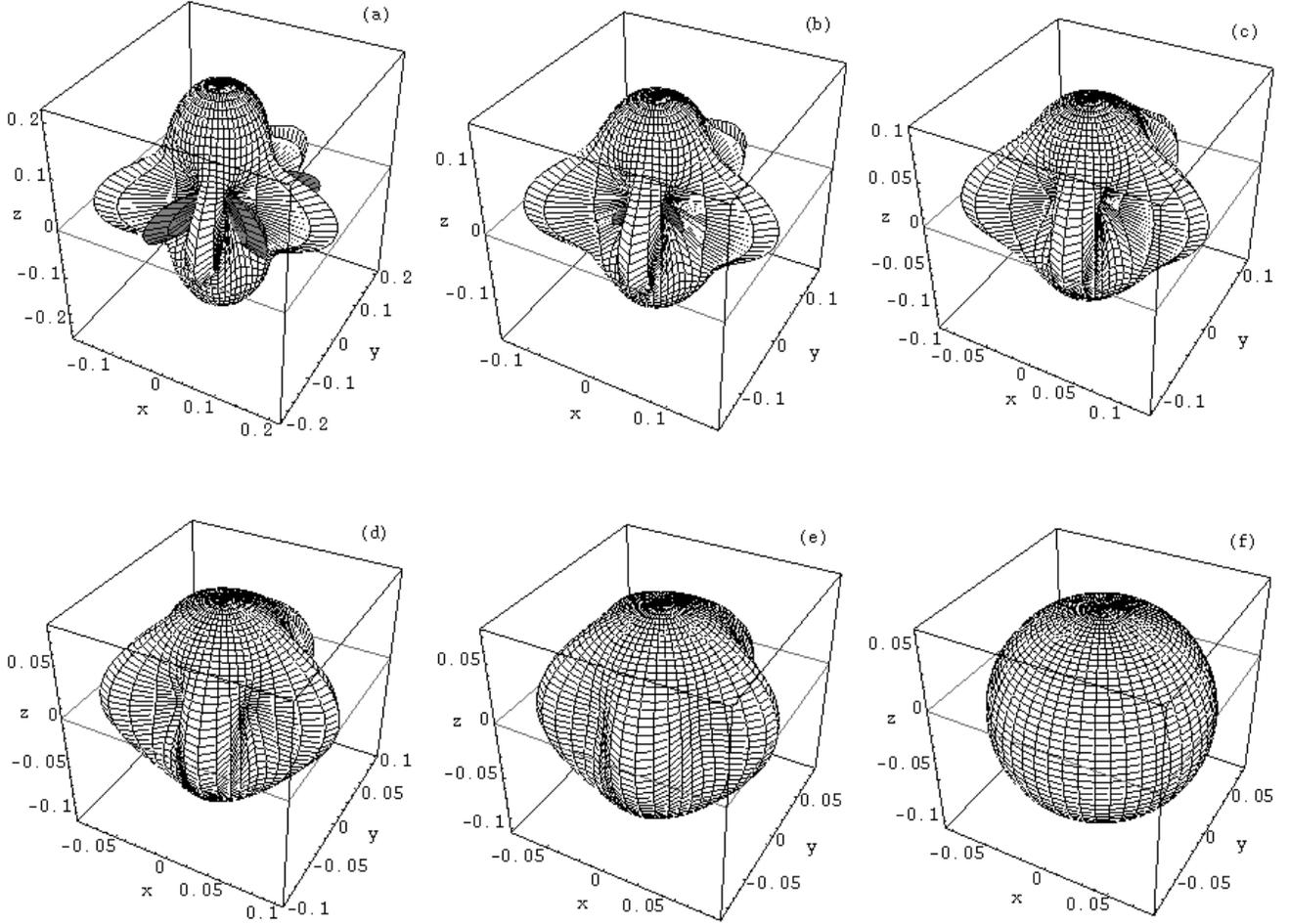}
\caption{
Polar plots of the temporal change of the Wigner function representing the 
decoherence and
dissipation of the initial polar cat state (\ref{SC0}), made of 5 atoms and 
shown in Fig.~\ref{fig:polcat}, for $\langle n \rangle=10$.
The dynamics of the corresponding density matrix elements is shown in 
Fig.~\ref{fig:dmplot}(c). 
The time instants are the following (in units of $1/\gamma$): 
(a) 0.01, (b) 0.019 ($=t_{\text{dec}}$), (c) 0.031 ($=t_{\text{ncl}}$), 
(d) 0.045, (e) 0.058 ($=t_{\text{diss}}$), (f) 0.25. 
}
\label{fig:wfN5n10}
\end{figure}

Figures \ref{fig:wfN5n0} and \ref{fig:wfN5n10} show the polar plots of 
the spherical Wigner function transforming its shape in time 
at $\langle n \rangle = 0$ and $\langle n \rangle = 10$,
respectively. The initial state is a polar cat made of $N=5$
atoms and its Wigner function is shown in Fig.~\ref{fig:polcat}.

%
%

In Figures \ref{fig:wfN5n0} and \ref{fig:wfN5n10} the following main 
characteristics of the
process can be identified. The decoherence is shown by the 
decreasing and finally disappearing ripples along the equator.
The vanishing of non-classicality, i.e. the decrease of the parameter
$\nu$, can be easily recognized as the decrease of the negative (dark)
wings. At nonzero temperature they disappear exactly at $t_{\text{ncl}}$,
as shown in Fig. \ref{fig:wfN5n10}(c).
The dissipation is represented by the approach of the initial upper and 
lower bumps to each other. 
At zero temperature (Fig. \ref{fig:wfN5n0})  the upper bump disappears 
and goes over to the lower one.  This stationary shape of the Wigner 
function corresponds to the lowest coherent state $|j,-j\rangle$ \cite{DAS}.
For $\langle n \rangle = 10$, when 
the stationary energy is close to the initial energy, not only the upper
bump moves downwards but also the lower one lifts upwards. The stationary
Wigner function has nearly a spherical symmetry, although its center 
is not in the origin.

In agreement with Fig. \ref{fig:cht}, the plots of Figures \ref{fig:wfN5n0} 
and \ref{fig:wfN5n10} show, that the timescales of the decoherence and
of the dissipation are very close to each other in the case of 5 atoms, 
for zero temperature they are practically the same. Then the spherical 
Wigner function exhibits considerable azimuthal modulation (ripples) also 
at $t_{\text{diss}}$.

We may come back finally to the question of finding the characteristic time 
of non-classicality $t_{\text{ncl}}$.      
According to the arguments given after Eq. (\ref{rhotrf}) it is sufficient
to study the Wigner function within a  $\phi$-range
of length $2\pi/N$, e.g. $0\leq \phi \leq 2\pi/N$, because it is invariant
with respect of rotations by $\phi=k {2\pi\over N},\ (k=1,2, \dots, N) $
i.e. it has $C_{N}$ symmetry at all times. Therefore the spherical Wigner 
function 
of a polar cat state, while subject to dissipation and decoherence, has its
minimum value always at $\phi = \pi/N$. 
Thus in order to calculate $t_{\text{ncl}}$, 
it is sufficient to follow the
time evolution of the section $W(\theta,\phi=\pi/N)$.
Further, in connection with the calculation of the measure of 
non-classicality $\nu$, it is sufficient to consider the above mentioned 
$\phi$-range when evaluating the integrals $I_+$ and $I_-$.

%
%
%
%

\section{Conclusions}

We have considered a class of states in an ensemble of two-level atoms, 
a superposition of two distinct atomic coherent states which are called 
atomic Schr\"{o}dinger cat states.
According to the relative positions of the constituents we have defined 
polar and nonpolar cat states.  We have investigated their properties 
based on the spherical Wigner function, which has been proven to be a 
convenient tool to investigate the quantum interference effects. 

We have shown that nonpolar cat states generally exhibit squeezing,
for which we have introduced the measure ${\cal S}$. 
The squeezing depends on the separation of the components 
of the cat and on the number of the atoms the cat is consisting of. 
By solving the master equation of this system embedded in an external 
environment we have determined the characteristic times of decoherence, 
dissipation and non-classicality of an initial polar cat state. 
We have shown how these depend on the number of the microscopic
elements the cat consists of, and on the temperature of the environment.
Our results show that the decoherence of the polar cat state is 
surprisingly slow: $t_{\text{diss}}/t_{\text{dec}}$ 
is less then a half of an order of magnitude for zero temperature,
making these states potentially significant in several areas of quantum
physics, e.g. experimental studies of decoherence, quantum computing and 
cryptography. 
We have visualised the process, governed by the interaction with 
the external environment, using the spherical Wigner function.
Its transformation in time reflects the characteristics of the behaviour
of the atomic subsystem in a suggestive way.

%
%
%
%

\section*{Acknowledgments}

The authors thank L.~Di\'{o}si, T.~Geszti, F.~Haake, J.~Janszky and 
W.~P.~Schleich for enlightening discussions on the subject, 
and Cs.~Benedek for his help in figure plotting.
One of the authors, A.~C. is grateful to the DAAD for financial
support. 
This work was supported by the Hungarian Scientific Research Fund 
(OTKA) under contracts T022281, F023336 and M028418.

%
%
%
%

\end{document}